# Minimax-Regret Climate Policy with Deep Uncertainty in Climate Modeling and Intergenerational Discounting

Stephen J. DeCanio[1], Charles F. Manski[2], and Alan H. Sanstad[3]

[1]University of California, Santa Barbara, Emeritus
[2]Department of Economics and Institute for Policy Research, Northwestern University
[3]Berkeley, California

## Abstract

Integrated assessment models have become the primary tools for comparing climate policies that seek to reduce greenhouse gas emissions. Policy comparisons have often been performed by considering a planner who seeks to make optimal trade-offs between the costs of carbon abatement and the economic damages from climate change. The planning problem has been formalized as one of optimal control, the objective being to minimize the total costs of abatement and damages over a time horizon. Studying climate policy as a control problem presumes that a planner knows enough to make optimization feasible, but physical and economic uncertainties abound. Manski, Sanstad, and DeCanio (2021) proposed and studied use of the *minimax-regret* (MMR) decision criterion to account for deep uncertainty in climate modeling. Here we study choice of climate policy that minimizes maximum regret with deep uncertainty regarding both the correct climate model and the appropriate time discount rate to use in intergenerational assessment of policy consequences. The analysis specifies a range of discount rates to express both empirical and normative uncertainty about the appropriate rate. The findings regarding climate policy are novel and informative. The MMR analysis points to use of a relatively low discount rate of 0.02 for climate policy. The MMR decision rule keeps the maximum future temperature increase below 2°C above the 1900-10 level for most of the parameter values used to weight costs and damages

**Acknowledgments:** We appreciate helpful comments from Valentyn Litvin and Robert Pindyck.

.

# 1. Introduction

Analysis of *integrated assessment* (IA) models enables quantitative evaluation of the benefits and costs of alternative climate policies. IA models are long-run (century-scale or more) descriptions of the global economy including the energy system and its role in economic production. These models incorporate representations of the climate and the links between the climatic effects of greenhouse gas (GHG) emissions and their impacts on the economy. IA models have become the primary tools for comparing policies that seek to reduce GHG emissions.

Policy comparisons have often been performed by considering a social planner who seeks to make optimal trade-offs between the costs of carbon abatement and the economic damages from climate change, at a global scale. The planning problem has been formalized as an optimal-control problem with three key components: (1) equations coupling GHG emissions and abatement to the accumulation of GHGs in the atmosphere and resulting temperature increases; (2) a damage function that quantifies economic effects of climate change in terms of the loss of global economic output as a function of temperature increases; and (3) an abatement cost function that expresses the cost of actions to reduce GHG emissions relative to a stipulated baseline emissions trajectory. Costs and damages at a point in time are expressed in terms of reductions in gross world product at that time. The control problem is to minimize the total costs of abatement and damages over a time horizon.

Studying climate policy as an optimal-control problem presumes that a planner knows enough about the global climate and economic systems to make optimization feasible. However, uncertainties abound. Physical and economic uncertainties have been handled in different ways.

The physical scientists whose research informs component (1) of IA models have performed multi-model ensemble (MME) analysis (Taylor et al., 2012). In the absence of a consensus climate model, they have developed multiple distinct models. To cope with inter-model *structural uncertainty*, they compute simple or weighted averages of the outputs of MMEs. However, choosing appropriate weights has been problematic.

The economists whose research informs components (2) and (3) have estimated multiple damage functions and abatement cost functions. In general, economists have not performed MME analyses that combine multiple functions by weighted averaging. They have instead reported disparate findings, stemming from their separate studies.

Manski, Sanstad, and DeCanio (2021) (M-S-D hereafter) framed structural uncertainty in climate modeling as a problem of *partial identification*, generating *deep uncertainty*. This problem refers to situations in which the underlying mechanisms, dynamics, or laws governing a system are not completely known and cannot be credibly modeled definitively even in the absence of data limitations in a statistical sense. M-S-D proposed use of the *minimax-regret* (MMR) decision criterion to account for deep climate uncertainty in integrated assessment without weighting climate model forecasts. They developed a theoretical framework for cost-benefit analysis of climate policy based on MMR and applied it computationally with a simple illustrative IA model.

To simplify the computational analysis, M-S-D studied MMR decision making in the presence only of physical-science uncertainty regarding the correct climate model. We specified damage and cost functions with functional forms and parameter values found in the literature on IA models. We engaged economic uncertainty only by exploring the sensitivity of findings to the specified parameter values of the damage and cost functions.

To move from illustrative analysis towards realistic comparison of climate policies, it is important to recognize joint deep uncertainty in the physical and economic components of IA models. Among the many economic aspects of IA models that have lacked consensus, perhaps the most contentious has been how a planner should assess the costs and benefits of policies across future generations. In this paper, we study choice of climate policy that minimizes maximum regret with deep uncertainty regarding both the correct climate model and the appropriate intergenerational assessment of policy consequences.

Economists have long framed intergenerational policy assessment using a time discount rate. They have evaluated climate policies by the present discounted value of the sum of abatement costs and the corresponding damages. However, there has long been debate about what discount rate to use; see, for example, Arrow et al. (2014) and Heal and Milner (2014). The choice is highly consequential. Low discount rates favor policies that reduce GHG emissions aggressively and rapidly (Emmerling et al., 2019). High rates favor policies that act more modestly and slowly. To express deep uncertainty, we suppose that the appropriate discount rate lies within an interval that covers the spectrum of rates that have been used in the literature. We suggest that consideration of this range of discount rates may be an attractive way for a planner to cope with normative uncertainty about the appropriate rate.

From a mathematical perspective, the computational analysis in this paper is a straightforward generalization of the analysis in M-S-D. That work supposed that the correct physical climate model is one of six prominent models in the literature on climate science, whereas the correct economic model is known. Given uncertainty about the climate model, M-S-D supposed that a planner compares six policies, each of which chooses an emissions abatement path that is optimal under one and only one of the six climate models. Regret is the loss in welfare if the model used in policy making is not correct and, consequently, the chosen abatement path is actually sub-optimal. The MMR rule chooses a policy that minimizes the maximum regret, or largest degree of sub-optimality, across all six climate models.

Here we also suppose that the correct climate model is one of the six models examined in M-S-D. We characterize uncertainty about the discount rate by supposing that it takes one of the seven values $\{0.01, 0.02, \ldots, 0.07\}$, a range that covers the rates commonly used. As we explain later, this range reflects both empirical uncertainty about the future of the economy and normative uncertainty (or perhaps disagreement) about how the current population values the welfare of future generations.

Given joint uncertainty about the climate model and the discount rate, we suppose that a planner compares forty-three policies. Each of forty-two policies chooses an emissions abatement path that is optimal under one of the six climate models and seven discount rates. The remaining one is the benchmark of a passive policy in which the planner chooses no abatement. With this setup, there are forty-two {discount rate, model} pairs, any of which is possibly correct. The regret of a specified policy under each pair is the loss in welfare if its abatement path is sub-optimal. The MMR criterion chooses a policy that minimizes maximum regret across all forty-two {discount rate, model} pairs.

Although the mathematical generalization of the earlier analysis is straightforward, the substantive findings regarding climate policy are novel and informative. The MMR analysis points to use of a relatively low discount rate for climate policy. The MMR decision rule keeps the maximum future temperature increase below 2℃ for most of the parameter values used to weight costs and damages.

In what follows, Section 2 describes how the physical-science and economics literatures have sought to cope with uncertainty about the correct climate model and discount rate respectively. Section 3 formalizes MMR policy choice, generalizing the IA model of M-S-D to incorporate discount-rate uncertainty. Section 4 presents our computational model and Section 5 gives the findings. Section 6 discusses the contributions and limitations of this work.

## 2. Prevalent Approaches to Climate and Discount-Rate Uncertainty

### 2.1. Averaging Outputs of MMEs of Climate Models

The climate is a complex system comprising many different physical processes occurring at a range of spatial and temporal scales, which climate models aim to represent in a tractable manner. All climate models are based on a specific set of deterministic nonlinear partial differential equations describing large-scale atmospheric dynamics. However, implementation of the equations in particular models is subject to numerous practical choices involving discretization, solution methods, and other details. Moreover, other components of the system – such as cloud formation and heat transfer between land surfaces and the atmosphere – are not yet fully understood and must be approximated. For these reasons, multiple climate models have been developed and are currently in use, each reflecting different but credible choices in model design and implementation. Existing models yield different projections of the global climate. Neither a "consensus" climate model nor definitive quantitative climate projections can be specified with current knowledge (Pindyck, 2022). The range of projections produced by different climate models is a gauge of deep uncertainty about the climate system given the current state-of-the-science.

Virtually all methods of MME analysis combine model outputs into single projections of future climate variables. A primary reason is that modelers have perceived policymakers as requiring single projections (as functions of particular GHG emissions scenarios) for use in decision-making (Parker 2006). However, climate researchers have recognized persistent methodological problems in combining model projections (Tebaldi and Knutti, 2007; Sanderson, 2018).

A common technique is to take the simple average across model projections of policy-relevant variables such as increases in global mean temperature due to anthropogenic carbon emissions. But computation of simple averages of predictions assumes that equal weight should be given to each model, an assumption lacking a compelling foundation (Knutti, 2010). Hence, researchers may instead compute weighted average projections when they believe that models can be ranked with respect to relative accuracy. However, model performance with respect to specific variables in historical data has not been demonstrated to imply skill in predicting climate (Flato et al., 2013), weakening the case for this approach to weighting projections for policy applications.

Combining climate model ensemble outputs into single projected trajectories of the future global climate remains a challenging and unresolved problem. As summarized in the recent Intergovernmental Panel on Climate Change (IPCC) physical sciences report, "…despite some progress, no universal, robust method for weighting a multi-model projection ensemble is available…" (Lee et al., 2021). This state of affairs poses a quandary for policymakers who rely on climate model output to formulate strategies for GHG emissions abatement and other approaches to address climate change.

2.2. Uncertainties and Disagreements Regarding the Discount Rate

IA models are subject to uncertainty in their economic assumptions as well as in their representation of the climate (Heal and Miller, 2014; Weyant, 2017). The paradigmatic example is Nordhaus's DICE (Dynamic Integrated Climate Economy) model, the most influential IA model of the last several decades (Nordhaus, 2019). In DICE and similar models, the economic losses from climate change are represented by damage functions that give the decreases in world-wide output resulting from increases in mean global temperature, as a proportional reduction or in dollar terms. These functions have uncertain theoretical and empirical grounding (Pindyck, 2013).

Economists study dynamic optimization by a social planner, which entails discounting to quantify the present value of future economic costs and benefits. The appropriate definition and magnitude of the discount rate is a long-standing and contentious issue in climate change economics and integrated assessment modeling (e.g., Ackerman et al., 2009; Arrow et al., 2014; Dasgupta, 2019; Pindyck, 2017; Weisbach and Sunstein, 2017). Controversy persists in part due to the fact that choice of an appropriate discount rate is not only an empirical question regarding the future of the economy. It is also a normative matter of ethics, concerning social preferences for equity across future generations which vary in their time of existence and in their levels of consumption (Dasgupta, 2008).

A simple version of the famous *Ramsey formula* (Ramsey, 1928) provides a transparent expression of the interplay of ethical and empirical considerations in choosing a discount rate. Paraphrasing the exposition in Arrow et al. (2014), let the planner's utilitarian social welfare function be additively separable in the utility of future generations. Let $\rho$ be the rate at which the social planner discounts the utility of future generations. Let the utility of a representative consumer be an increasing and concave function of consumption, with constant elasticity $(-\eta)$ of marginal utility with respect to consumption. Let $g_t$ be the annualized growth rate of consumption between time 0 and a future time $t$. Ramsey showed that it is optimal to discount future consumption between the present (time 0) and time $t$ at the rate

$$(1) \qquad\qquad \delta_t \;\equiv\; \rho + \eta\, g_t\,.$$

Of the variables on the right-hand side, $g_t$ describes future consumption growth in the economy. From the perspective of the present, the empirical value of $g_t$ may be uncertain, perhaps deeply so. Such uncertainty is similar conceptually to the uncertainty that climate modelers face as they attempt to project the future trajectory of climate variables.

The time-invariant quantities $\rho$ and $\eta$ are normative parameters. The value of $\rho$ formalizes how the planner views intergenerational equity, with $\rho = 0$ if the planner gives equal weight to the welfare of all future generations and $\rho > 0$ if the planner weights welfare more heavily in the near future than in the distant future. The value of $\eta$ formalizes how the planner views the desirability of consumption equity. Under conventional utilitarian presumptions, the marginal utility of consumption decreases as consumption increases. Therefore, a larger value of $\eta$ combined with positive $g_t$ (i.e., future generations are richer) implies that the planner should use a larger discount rate to evaluate costs and benefits, as expressed in equation (1).

Being normative parameters, $\rho$ and $\eta$ are not subject to empirical uncertainty in the sense of $g_t$ or climate projections. Nevertheless, a social planner may feel normative uncertainty about what values are appropriate to use. Supposing that the planner aims to represent society, a source of this uncertainty may be normative disagreements within the present population. Such disagreements were evident, for example, in a highly public dispute between Nordhaus (2007), whose policy analysis used the value $\rho = 0.03$, and Stern (2006), whose analysis used the value $\rho = 0.001$. This difference was highly consequential. Stern concluded that policy should seek to reduce GHG emissions aggressively and rapidly. Nordhaus favored policies that act more modestly and slowly.

Recognizing that conclusions about climate policy may depend critically on the discount rate used, economists have struggled to do more than debate the issue. Weitzman (2001) suggested use of a weighted average of the discount rates considered in the climate-economics literature, a procedure akin to the weighted averaging performed by climate scientists in MME analysis. Heal and Milner (2014) mention other possible ways to obtain a discount rate that a planner might find appropriate to use.

We argue against any attempt to cope with empirical and normative uncertainty by choosing a single discount rate. Instead, we study formation of climate policy recognizing a set of possibly appropriate discount rates. The remainder of this paper shows how.

## 3. Minimax-Regret Policy Evaluation

We study a straightforward extension of the MMR policy-choice problem posed by M-S-D. To begin, we specify the optimal-control problem that a planner would solve in the absence of uncertainty.

### 3.1. The Optimal-Control Problem

Let $B_t$ represent baseline GHG emissions at time $t$, $A_t$ be GHG abatement or mitigation actions at time $t$ under some climate policy, measured in the same units as emissions, $C(A_t)$ be the cost of these actions, and $E_t^{A_t} = B_t - A_t$ be the resulting net emissions. (The terms abatement and mitigation are both used in the literature to describe actions reducing GHGs and related measures.) We refer to $A_t$ and $E_t^{A_t}$ as "paths" or "trajectories," and we assume that abatement paths are chosen from some space of feasible paths.

Emissions paths are used as inputs to a climate model $M$. We focus on the global mean temperatures projected by $M$ as a function of these paths. Thus, let $T(E_t^{A_t}, M)$ be the global mean temperature at time $t$ determined by the GHG trajectory $E_t^{A_t}$ when it is simulated in the climate model $M$. Then a damage function, as discussed above, can be written as $D\left(T\left(E_t^{A_t}, M\right)\right)$.

For an abatement path $A_t$ and climate model $M$, denote the associated total cost (abatement plus damages) at time $t$ as

$$(2) \qquad \mathbb{C}(A_t, M) \equiv C(A_t) + D\left(T\left(E_t^{A_t}, M\right)\right).$$

A policymaker seeks to minimize the present value of cumulative cost over a planning horizon which, as is customary in the climate economics literature, we assume to be infinite. The optimal control problem given a particular climate model $M$ is to solve

$$(3) \qquad \min_{A_t} \int_0^\infty \mathbb{C}(A_t, M) e^{-\delta t} dt \, ,$$

where $\delta$ is a time-invariant discount rate. In this approach, the optimal $A_t$ is chosen with commitment at time zero – that is, it is not updated over time as new climate or cost information is obtained. As stated, (3) is a deterministic optimization problem that, under certain technical assumptions regarding the feasible abatement path space and the cost and damage functions, has a unique solution. We will assume that such conditions hold for the series of problems we describe.

## 3.2. The Minimax-Regret Decision Rule

Now let $\boldsymbol{\Delta} = \{\delta_1, \ldots, \delta_K\}$ be a set of possibly appropriate discount rates and $\mathbf{M} = \{M_1, \ldots, M_N\}$ be a model ensemble. The planner now faces the problem of minimizing total present-value cost over the infinite horizon while recognizing joint {discount rate, model} uncertainty. For a particular discount rate $\delta_i$ and model $M_j$, let $A^*_{t;\delta_i,M_j}$ be the abatement path defined by

$$(4) \qquad A^*_{t;\delta_i,M_j} = \arg\min_{A_t} \int_0^\infty \mathbb{C}(A_t, M_j) e^{-\delta_i t} \, dt$$

That is, this cost-minimizing $A^*$ is the *optimal* trajectory when the discount rate is $\delta_i$ and the model is $M_j$. Let $\mathbb{C}^*\left(A^*_{t;\delta_i,M_j}, \delta_i, M_j\right)$ be the associated minimum cost:

$$(5) \qquad \mathbb{C}^*\left(A^*_{t;\delta_i,M_j}, \delta_i, M_j\right) = \int_0^\infty \mathbb{C}(A^*_t, M_j) e^{-\delta_i t} dt$$

(Note the change in notation: Previously, $\mathbb{C}(A_t, M_j)$ was total cost at time $t$; now, $\mathbb{C}^*\left(A^*_{t;\delta_i,M_j}, \delta_i, M_j\right)$ is total discounted cost, i.e., an integral.)

Now consider any feasible abatement trajectory $A_t$. The *regret* $\mathbb{R}(A_t, \delta_i, M_j)$ associated with $A_t$, when discount rate $\delta_i$ and climate model $M_j$ describe the actual state of the world, is the difference between the cost of $A_t$ in the actual state of the world and the cost of the *optimal* policy associated with $\delta_i$ and $M_j$:

$$(6) \qquad \mathbb{R}(A_t, \delta_i, M_j) = \int_0^\infty \mathbb{C}(A_t, M_j) e^{-\delta_i t} dt - \mathbb{C}^*\left(A^*_{t;\delta_i M_j}, \delta_i, M_j\right).$$

To apply the MMR rule, the planner first considers each feasible abatement path $A_t$ and finds the model and discount rate combination that maximizes regret as defined in Equation (6), solving the problem

$$(7) \qquad \max_{\delta_i, M_j} \ \mathbb{R}(A_t, \delta_i, M_i) = \max_{\delta_i, M_j} \left[ \int_0^\infty \mathbb{C}(A_t, M_j) e^{-\delta_i t} dt - \mathbb{C}^*(A_{t;\delta_i,M_j}^*, \delta_i, M_j) \right] .$$

The MMR solution is then to find $A_t$ to solve the problem

$$(8) \qquad \min_{A_t} \left[ \max_{\delta_i, M_j} \ \mathbb{R}(A_t, \delta_i, M_j) \right] .$$

### 3.3. Use of $\Delta$ to Express Empirical and Normative Uncertainty

In research on decision making under uncertainty, the term "uncertainty" has usually referred to incomplete knowledge of the empirical environment of the decision maker, commonly called the "state of nature" or the "state of the world." In study of climate policy, this interpretation of uncertainty applies to incomplete knowledge of the future global temperature, abatement costs, and damages that will occur if alternative climate policies are chosen. It also applies to uncertainty about the discount rate that stems from difficulty in predicting the future of the economy.

For example, economists using the Ramsey formula to specify a discount rate have studied policy formation when the growth path $g_t$ of future consumption is generated by an assumed stochastic process; see the discussion in Arrow et al. (2014). A planner may feel unable to specify a credible stochastic-process for $g_t$, so uncertainty about the discount rate is deep. If so, the MMR rule given in (8) provides a reasonable approach to policy making with the Ramsey formula.

In the computational analysis of Section 5, we will use the MMR rule in (8) to embrace normative as well as empirical uncertainty about the appropriate discount rate to use when evaluating climate policy. We need to consider normative uncertainty (or perhaps disagreement) because, as discussed in Section 2.2, debate among economists about the appropriate discount rate has stemmed from more than empirical uncertainty about the future economy. Notably, the dispute between Nordhaus (2007) and Stern (2006) regarding optimal climate policy, mentioned earlier, occurred largely because they specified different values for the normative parameter $\rho$ in the Ramsey formula.

Our use of the set $\mathbf{\Delta}$ to express both empirical and normative uncertainty regarding the appropriate discount rate departs only modestly from the usual decision-theoretic focus on empirical uncertainty if the planner is a utilitarian entity who has incomplete knowledge of the intergenerational preferences of the present population. Then the planner's normative uncertainty has an empirical source, namely incomplete knowledge of the population preferences that a utilitarian would seek to maximize. Pushing this idea further, the planner may face the difficult task of representing a population whose members may not themselves be clear about their pure time preferences or willingness to accept intergenerational inequalities.

Social planning using $\mathbf{\Delta}$ to express normative uncertainty is a more radical departure from the decision-theoretic norm if the underlying problem is clear and yet sharp normative disagreements exist within the present population. That is, a segment of the population may strongly value intergenerational equity whereas another segment may be less concerned with the fate of future generations. In this case, one may think it necessary to abandon the idealization of a utilitarian planner and replace it with conceptualization of policy making as a non-cooperative political game.

We nonetheless find it attractive to study MMR decision making even in this challenging setting. The reason is that the MMR rule has some appeal as a broadly acceptable mechanism for policy choice. Recall that the *regret* of a policy in a specified state of nature measures its degree of sub-optimality in that state, and that maximum regret measures the maximum degree of sub-optimality across all states. Suppose that the members of a heterogeneous present population disagree on what {discount rate, model} should be considered the "true" state of nature. Then use of the MMR rule to choose policy minimizes the maximum degree of sub-optimality that will be experienced across the population.

The notion that the MMR policy may be broadly acceptable because it minimizes maximum sub-optimality is reminiscent of Rawls's consideration of social decision making behind a veil of ignorance. As with Rawls, we find it appealing to minimize some measure of the maximum harm experienced by a heterogeneous population. However, the two settings differ in important respects. Formation of intergenerational climate policy differs from Rawls's concern with static income distribution, and the MMR decision rule differs from Rawls's consideration of maximin decisions.

## 4. Computational Model

A number of elaborate computational IA models have been proposed in the literature and are used in climate policy analysis. To show in broad terms the consequences of adoption of the MMR decision rule in this context, we instead present a very simple IA model that summarizes the essential economic and physical mechanisms at work. While the standard in the IA literature is to report results only about a century into the future, analyzing the uncertainty associated with discount rates necessitates attention to longer time horizons because phenomena in the more distant future that are negligible in economic terms with conventional discounting become salient with low rates. Notwithstanding the speculative nature of extreme emissions reduction or reversal scenarios (see Appendix A for additional discussion), we follow the convention of assuming them in our computational model, both in a baseline emissions trajectory and in abatement paths.

### 4.1 Model Details

To illustrate quantitatively the solution of MMR equation (8) in Section 3.2, we specify functional forms and parameters for the climate damages, abatement costs, and climate dynamics in order to create a simple IA model. To define simple reduced forms of complex climate model dynamics, we draw on the work of Matthews et al. (2009). They showed that the "carbon-climate response" (CCR), the change in global mean temperature over periods of decades or longer, varies approximately linearly with the increase in cumulative carbon emissions over the same period. We define *net* cumulative emissions as

(9)
$$\mathbf{E}_t^{A_t} = \int_0^t E_t^{A_t} dt = \int_0^t (B_t - A_t) dt$$

Matthews et al. analyzed the case $A_t = 0$; that is, no abatement relative to the baseline at any time. Note that there is no requirement our definition that $(B_t - A_t)$ be non-negative. $A_t$ exceeding $B_t$ implies adoption of mitigation measures that yield negative net emissions at some time $t$. Doing so may result in declining net cumulative emissions if sustained long enough.

The CCR parameter $m(M_i)$, or $m_i$ for short, associated with each full-scale numerical climate model is estimated by determining the model's projected temperature response when it is driven by a carbon emissions path according to

(10)
$$T_i = m_i \mathbf{E}_t^{A_t}, \qquad i = 1, \ldots .6$$

where $T$ henceforth indicates the temperature increase over its initial value at time $t = 0$. The CCRs vary across climate models, reflecting structural and other forms of uncertainty. The CCR allows incorporating both reduced-form climate dynamics and deep climate model uncertainty into our simple IA modeling framework.

Our model ensemble $\mathbf{M}$ is obtained by drawing on results from simulations of six "Earth System Models (ESMs)," which combine physical climate models with representations of biogeochemistry in order to simulate the complete atmospheric, oceanic, and terrestrial carbon cycle. These ESMs were used in the Climate Model Intercomparison Project Phase 5 (CMIP5), a study under the auspices of the World Climate Research Programme. We estimate CCR parameters $m_i$ using historic and projected emissions and temperature data from each of the six ESMs (see M-S-D for details). Our model ensemble can then be described succinctly by $\mathbf{M} = \{m_1, \ldots, m_6\}$. The models and their associated $m_i$ are shown in Table 1.

| Table 1<br>**Earth system models used to estimate Carbon-Climate Response (CCR)<br>parameters, with estimated CCR values (°C per teraton carbon)** | |
|---|---|
| ***Model and model number*** | ***CCR*** |
| 1. GFDL-ESM-2G - Geophysical Fluid Dynamics Laboratory Earth System Model version 2G | 0.00157 |
| 2. BCC-CSM-1 - Beijing Climate Center Climate System Model version 1.1 | 0.00186 |
| 3. FIO-ESM - FIO-ESM - First Institute of Oceanography Earth System Model | 0.00194 |
| 4. Had-GEM2-ES - Hadley Global Environmental Model 2 - Earth System | 0.00229 |
| 5. IPSL-CM5A-MR - Institut Pierre Simon Laplace Coupled Model 5A - Medium Resolution | 0.00236 |
| 6. MIROC-ESM - Model for Interdisciplinary Research on Climate - Earth System Model | 0.00244 |
| Taylor et al. (2012), Collins et al. (2013) | |

Next, we specify abatement cost and climate damage functions in quadratic form to implement the IA model as an optimal control problem, allowing for plausible non-linearity in these functions as the abatement effort and the temperature increase:

(11) $$C(A_t) = \tfrac{1}{2}\,\alpha A_t^2$$

(12) $$D(T_t) = \tfrac{1}{2}\,\beta T_t^2$$

where $\alpha$ and $\beta$ are weighting parameters calibrated to numerical estimates in the climate economics literature, and $T_t$ is, as above, global mean temperature increase as of time $t$. For damages, the quadratic form and the value of $\beta$ are taken from a statistical survey by Nordhaus and Moffat (2017), and comprise these researchers' preferred regression model for approximating existing empirical damage estimates. The quadratic form and value of $\alpha$ are derived from Dietz's and Venman's (2019) synthesis of global marginal abatement costs reported in the Intergovernmental Panel on Climate Change's Fifth Assessment Report (Clarke et al., 2014).

A baseline emissions trajectory $B_t$ is derived from the so-called "Representative Concentration Pathway (RCP) 8.5" scenario in its extended version to year 2500, which envisions a relatively high growth rate of global carbon emissions from fossil fuel use through the 21st century, followed by a peak or plateau period of constant emissions until 2150, and then a decline to a very low level by 2250 (Riahi et al., 2011, Meinshausen et al., 2011). In its original form with a 2100 time horizon, the RCP 8.5 reflects an absence of explicit global climate policy. This, and several other extended RCP scenarios assuming different emissions paths, were devised for research purposes, with no explanation given of the policies, technological advances, or other factors that could bring about this peaking, leveling, and decline. See Appendix A for discussion of issues in specifying a realistic baseline trajectory for IA modeling of climate policy.

A functional form having the same general shape as the RCP 8.5 was fitted by nonlinear least squares using the Levenberg-Marquardt method in Mathematica (2019). The fitted equation for $B_t$ is

$$(13) \qquad B_t = \left( \theta t + \frac{B_0}{\exp\,(\theta\varphi)} \right) \theta \, \exp\,\left( -(t - \varphi) \right).$$

This equation smooths connected segments of the extended RCP 8.5, including the plateau during the first half of the 22nd century, and captures its rapid 21st-century increase and subsequent dramatic decline. Equation (13) fits the scenario data well, with an $R^2$ of 0.927.

Combining these various components, the optimal control problem is to minimize, for a particular discount rate and model, the present value of abatement costs plus climate damages over an infinite horizon, subject to the dynamic relationship between cumulative emissions and temperature:

$$(14) \qquad \min_{A_t} \int_0^\infty \frac{1}{2} \left( \alpha A_t^2 + \beta T_t^2 \right) e^{-\delta t} dt$$

subject to

(15)
$$\frac{d}{dt}\mathbf{E}_t^{A_t} = E_t^{A_t} = B_t - A_t$$

(16)
$$T_t = m\mathbf{E}_t^{A_t}$$

(17)
$$\mathbf{E}_0^{A_t} = \mathbf{E}_0$$

where the last equation specifies an initial condition for net cumulative emissions. Applying standard solution techniques (see, e.g., Barro and Sala-i-Martin, 1995) yields first-order conditions including two coupled differential equations in abatement and the atmospheric greenhouse gas concentration associated with the optimal abatement:

(18)
$$\frac{dA_t}{dt} = \delta A_t - \frac{\beta m^2}{\alpha} \mathbf{E}_t^{A_t}$$

(19)
$$\frac{d\mathbf{E}^{A_t}}{dt} = B_t - A_t \quad .$$

These equations can be solved in closed form for the optimal abatement path $A_t$ and resulting optimal temperature, costs, damages, and present-value total cost.

We discussed our numerical reduced-form climate model ensemble $\mathbf{M}$ above. For numerical solution of the MMR problem in equation (8), it is also necessary to specify the discount rate set $\mathbf{\Delta}$. Following the discussion in Section 3.3, we pick seven possible discount rates ranging from a low of 0.01 to a high of 0.07 to represent the extent of empirically-based and normative uncertainty; these values roughly span the set of discounts rates that have been used in the climate economics literature. For numerical implementation, we pick the seven values {0.01, 0.02, 0.03, 0.04, 0.05, 0.06, 0.07} in this interval. It should be noted that a zero discount has also been analyzed and debated, but the optimal control problem stated in Equations (14) – (17) does not have a solution in this case because the transversality first-order condition is not satisfied. As an approximation, we selected $\delta = 0.01$ for the smallest potential $\delta$ value. The largest value, 0.07, corresponds to the real, pre-tax return on private investment (Arrow et al., 2013, citing U.S. Office of Management, 2003). The six reduced-form climate models in the ensemble $\mathbf{M}$ are defined by the different values of their CCRs as discussed above (see Table 1). With seven values in the ensemble $\mathbf{\Delta,}$ there are thus forty-two combinations of $\delta$ and $m$ representing the range of the two types of deep uncertainty we have discussed.

Before turning to uncertainty analysis, we illustrate how the model works, including the influence of the baseline emissions scenario approximating the RCP 8.5 and the implications of the discount rate uncertainty. The left-hand panel of Figure 1 shows the baseline $B_t$ and the optimal abatement $A_t$ for a particular set of parameters. The right-hand panel shows net cumulative emissions under $A_t$ and under a policy of no abatement ($A_t = 0$ for all $t$), respectively. Starting from the initial year, abatement increases along with baseline emissions, but with a lag: Through the 21$^{st}$ century the model does not find it optimal to abate baseline emissions completely. In the 22$^{nd}$ century, by contrast, optimal abatement begins to exceed the baseline, resulting in a decline in net cumulative emissions, as seen in the right-hand panel. This panel also shows that optimal abatement policy entails a significant reduction in net cumulative emissions relative to no abatement.

**Figure 1** — Trajectories of $B_t$, optimal $A_t$, $\mathbf{E}_t^{A_t}$, and $\mathbf{E}_t^{\text{noabatement}}$ for $m = 0.002286$, $\alpha = 0.000125$, $\beta = 0.018$, and $\delta = 0.05$

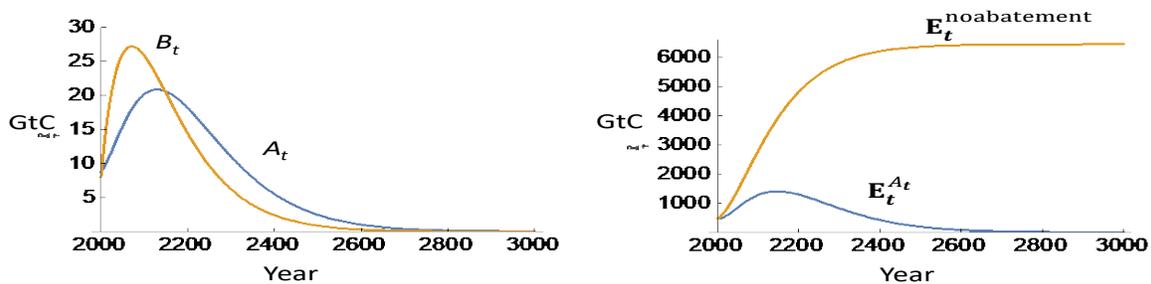

The computational model shows the importance of the discount rate. With the parameters of Figure 1, the *optimal* total economic loss from climate change, including abatement cost, is 0.51% of the present-discounted value of future global GDP. If the discount rate were 1%, this optimal value would be 3.51%.

By way of comparison, with no abatement and the parameters of Figure 1 (i.e., with a discount rate of 5%), the total economic cost of climate change would be 0.79% of the present-discounted value of future global GDP, while at a 1% discount rate these damages would be 42.02% of present-discounted value of future global GDP.

## 5. MMR Analysis

The computational model presented in Section 4 enables us to quantify the theoretical discussion in Section 3.2. For the middle values of the economic parameters $\alpha$ and $\beta$, the full set of regrets (from Equation (6)) across all possible pairs of $\delta$ and $m$ is shown in the three-part Table B-1 in Appendix B. Regrets can be calculated for any feasible abatement path $A_t$, but to keep the calculations tractable (and because a planner presumably would seek to pick a "best" policy) we focus on those paths that are optimal for those combinations of $\delta$ and $m$ that are in the $\mathbf{\Delta}$ and $\mathbf{M}$ ensembles. There are 42 such combinations. In Table B-1, a policy is defined by optimizing using the $\{\delta_i, m_j\}$ pair at the top of each column, while the $\{\delta_i, m_j\}$ values that correspond to the actual state of the world are indicated by the row headings on the leftmost border. For example, if policy is optimized assuming $\{0.03, m_2\}$, while the actual state-of-the-world is best described by $\{0.04, m_1\}$, the regret is 0.049, as highlighted in bold in the fourth row and tenth column of Part 1 of Table B-1. Appendix Figure B-1 gives a color-shaded plot of the 1849 numbers in Table B-1.

The maximum regret for each policy is given in the final row of the Table, across each of its three parts. From those maximum regrets, the minimum of them (i.e., Equation (8)) can be read off directly. For this ($\alpha$, $\beta$) combination it is 0.273, which appears at the bottom of the second column of Part 3.

To explore the sensitivity of the MMR to different combinations of the $\alpha$ and $\beta$ parameters, we calculated the full matrix of 1806 regrets, the maximum regret for each $\{\delta_i, m_j\}$ combination, and the MMR for nine combinations of $\alpha$ and $\beta$. The values of the MMR, along with the combination of $\delta$ and Model that yield the MMR, are given in Table 2. The most striking result exhibited in this Table is that, under all combinations of $\alpha$ and $\beta$, the discount rate corresponding to the MMR solution is 0.02, near the low end of the range of discount rates considered. Different climate models are picked by the MMR rule depending on the ($\alpha$, $\beta$) combination, although the three models appearing in Table 2 are the ones with the highest values of the CCR parameter $m$. However, in all cases the MMR decision rule points unambiguously to use of a relatively low discount rate in evaluating the costs and benefits of climate mitigation measures. The strong finding regarding the low value of $\delta$ selected by the MMR criterion is reinforced if we consider cases where the Model is known and only $\delta$ is uncertain. These results are not shown here, but are available from the authors on request.

The simple IA model also allows for calculation of the maximum temperature increase that will be reached for any policy path, and how long it will take to reach that temperature, two quantities of considerable importance in climate science and policy (Clarke et al. 2014, IPCC, 2021). Once the MMR policy combination of $\{\delta, m\}$ is selected, the abatement and net cumulative emissions paths are determined. The maximum temperature increase given such a path is determined by equation (10), and will occur when net cumulative emissions is at a maximum. The value of the CCR parameter in the actual state of the world may be different from that in the MMR policy combination.

| Table 2 | | | | | | | | |
| --- | --- | --- | --- | --- | --- | --- | --- | --- |
| **Values of MMR, uncertain Model and  $\delta$ , for combinations of  $\alpha$  and  $\beta$**<br>**Potential values of δ:  {0.01,0.02,0.02,0.03,0.04,0.05,0.06,0.07}** | | | | | | | | |
| *α = 0.000075*<br>*β = 0.014* | | | *α = 0.000075*<br>*β = 0.018* | | | *α = 0.000075*<br>*β = 0.022* | | |
| **Model** | **δ** | **MMR** | **Model** | **δ** | **MMR** | **Model** | **δ** | **MMR** |
| IPSL | 0.02 | **0.172** | HAD | 0.02 | **0.172** | HAD | 0.02 | **0.178** |

| *α = 0.000125*<br>*β = 0.014* | | | *α = 0.000125*<br>*β = 0.018* | | | *α = 0.000125*<br>*β = 0.022* | | |
| --- | --- | --- | --- | --- | --- | --- | --- | --- |
| **Model** | **δ** | **MMR** | **Model** | **δ** | **MMR** | **Model** | **δ** | **MMR** |
| MIROC | 0.02 | *0.266* | IPSL | 0.02 | *0.273* | IPSL | 0.02 | *0.284* |
| | | | | | | | | |
| *α = 0.0002*<br>*β = 0.014* | | | *α = 0.0002*<br>*β = 0.018* | | | *α = 0.0002*<br>*β = 0.022* | | |
| Model | δ | MMR | Model | δ | MMR | Model | δ | MMR |
| MIROC | 0.02 | *0.478* | MIROC | 0.02 | *0.436* | MIROC | 0.02 | *0.423* |

Because the actual state of the world is unknown, the temperature increase under the MMR policy cannot be known at the time the policy decision is made. What is known is that it will be less than or equal to the maximum over all six models, which will occur if MIROC is the true model because $m_6$ is the greatest of the CCRs. Table 3 shows these maxima for the nine combinations of $\{\alpha, \beta\}$, and the year at which it is reached. It can be seen from Table 3 that for almost all the parameter combinations considered here, the MMR policy keeps the peak temperature increase under 2℃. Only for the costliest abatement (cases with $\alpha = 0.0002$) does the maximum temperature increase exceed 2℃. Under the MMR abatement policy across the nine ($\alpha$, $\beta$) combinations, the global mean temperature will reach its maximum between 118 and 149 years from the initial point.  Although not shown in the Table, it is worth noting that without any abatement, the maximum temperature increase can be quite high, for example, it is 14.7°C as time goes to infinity for the middle values of $\alpha$ and $\beta$ and the Hadley model ($m_4$).

| Table 3 | | | | | | | | |
|---------|---|---|---|---|---|---|---|---|
| **Values of Maximum Temperature Increase (Tmax) and Years until it is reached, uncertain Model and $\delta$, for combinations of $\alpha$ and $\beta$** **Potential values of $\delta$:  {0.01, 0.02, 0.03, 0.04, 0.05, 0.06, 0.07}** | | | | | | | | |
| *$\alpha = 0.000075$* *$\beta = 0.014$* | | | *$\alpha = 0.000075$* *$\beta = 0.018$* | | | *$\alpha = 0.000075$* *$\beta = 0.022$* | | |
| **MMR Model** | **Years** | **Tmax** | **MMR Model** | **Years** | **Tmax** | **MMR Model** | **Years** | **Tmax** |
| IPSL | 124 | 1.248 | HAD | 121 | 1.055 | HAD | 118 | 0.877 |

| *$\alpha = 0.000125$* *$\beta = 0.014$* | | | *$\alpha = 0.000125$* *$\beta = 0.018$* | | | *$\alpha = 0.000125$* *$\beta = 0.022$* | | |
|---------|---|---|---|---|---|---|---|---|
| **MMR Model** | **Years** | **Tmax** | **MMR Model** | **Years** | **Tmax** | **MMR Model** | **Years** | **Tmax** |
| MIROC | 134 | 1.831 | IPSL | 130 | 1.564 | IPSL | 125 | 1.315 |
| | | | | | | | | |
| *$\alpha = 0.0002$* *$\beta = 0.014$* | | | *$\alpha = 0.0002$* *$\beta = 0.018$* | | | *$\alpha = 0.0002$* *$\beta = 0.022$* | | |
| **MMR Model** | **Years** | **Tmax** | **MMR Model** | **Years** | **Tmax** | **MMR Model** | **Years** | **Tmax** |
| MIROC | 149 | 2.660 | MIROC | 141 | 2.187 | MIROC | 135 | 1.859 |

## 6. Discussion

We consider the minimax regret rule to provide a reasonable way to form climate policy with both empirical uncertainty about the physical climate system and normative uncertainty regarding the discount rate. Our computational analysis of MMR decision making offers a new reason for using a low discount rate in climate policy analysis, a rate on the order of 2% per annum. This discount rate encompasses the pure rate of time preference, intergenerational inequality aversion, and projection of the economy's future rate of growth.

MMR decision making copes with deep uncertainty without adopting the extreme conservatism of minimax decisions. As discussed in Section 3.3, MMR enables a planner to deal with heterogeneous populations, who may not themselves be clear about their time preferences or willingness to accept intergenerational inequalities. There is no scientific or economic reason that everyone should hold the same normative values. Furthermore, some people may have only a vague understanding of discounting.

We also find it appealing to view the MMR decision rule as a consensus-building mechanism. Calculating regrets enables people with different values to see how implementation of alternative policies might play out from their perspectives. Having the appeal of limiting maximal harm across the population, choosing policy by MMR may be acceptable to holders of incompatible preferences.

Of course, the modeling in this paper does not address the philosophical problems that some may have with the utilitarian framework. Aggregating individualistic "utilities" may be a questionable basis for societal decisions. It is possible to aggregate monetary costs and benefits, but these can be translated into utilities only if Kaldor-Hicks compensations are implemented – something that rarely takes place in the real world. Some forms of compensation are impossible to accomplish: future generations cannot pay us to mitigate climate change.

The IA model described in our paper is simple and computationally tractable. This is partially because we have not considered all possible sources of uncertainty. As shown in Appendix A, the appropriate baseline emissions path is highly uncertain. Despite the best efforts of economists, the shapes and parameters of the abatement cost and climate damage functions are also uncertain. We have addressed this partially by sensitivity analysis, calculating MMR solutions with various parameters ($\alpha$, $\beta$) on abatement cost and climate damages. However, it would be better to expand our MMR analysis to encompass deep uncertainty about the correct values for these weights, a formidable computational task. Future research may narrow the realistic ranges of these parameters, as well clarify the shapes of the abatement cost and damage functions. Knowledge of realistic baseline scenarios may improve. Ongoing progress in physical climate modeling is likely to lead to better understanding of the relationship between greenhouse gas emissions, global temperatures, and other features of the geophysical system.

And yet, as discussed in Sections 2.2 and 3.3, there is an intrinsic difference between the discount rate, which depends on normative considerations, and the models/parameters of the physical climate and the economy. Geophysical and economic models may continue to be improved by new scientific research, but the normative parameters determining the discount rate involve issues that are beyond the reach of science. We suggest that the MMR approach to climate policy decision-making provides an attractive way to cope with both empirical and normative uncertainty.

**Appendix A.    Specifying the Baseline Emissions and Abatement Paths**

In Section 3.1 we defined a "net emissions" path as $E_t^{A_t} = B_t - A_t$, where $B_t$ is a baseline emissions path and $A_t$ is a path of carbon emissions abatement actions. In the climate economics literature, quantitative baselines have usually been specified on the basis of narrative scenarios describing in qualitative terms how the world economy, society, and energy system might evolve over the course of the twenty-first century and beyond. Projected abatement paths that reduce emissions relative to baselines are similarly conceived and quantified.

Both baseline scenarios and abatement paths have been developed to reflect a range of global GHG emissions projections, including both continued high fossil fuel use and dramatic reductions. Reductions are modeled as being achieved through measures including carbon pricing, emissions cap and trade systems, and adoption of low-carbon or non-carbon primary power generation – solar, wind, nuclear, and fossil with carbon capture and sequestration.  However, there are no generally accepted criteria stipulating what should be considered feasible or realistic in these low-carbon scenarios, whether technologically, economically, or politically. In many cases, scenarios project implementation of measures significantly beyond past or current levels. Indeed, there are examples in which scenarios project rapid global deployment of technologies that are at present still only in the development and demonstration phase.

Increasing attention is being directed toward even more radical measures, which would go beyond reducing emissions to reversing them.  "Carbon dioxide removal" (CDR) refers to human activities such as afforestation, ocean fertilization, and direct air capture and sequestration, which extract already-emitted carbon from the atmosphere and store it in underground reservoirs or other repositories. Aggressive deployment of CDR could result in "net negative" emissions, meaning that the rate of removal would exceed the rate of new emissions. But CDR on scales sufficiently large and sustained to meaningfully affect global climate change is at this point highly speculative, both in terms of technologically and with respect to its effects on the climate and other impacts, such as on agriculture. Nevertheless, it has come to the fore with the advent of highly ambitious global temperature change limits as an international policy goal. Notably, in modeling of emissions paths to prevent mean global warming of more than 1.5°C over the pre-industrial level, most scenarios in which this limit is achieved entail global net zero, and then negative, emissions during the twenty-first century, requiring CDR far beyond what is currently feasible, even on small, demonstration-level scales.

Another form of geoengineering is "Solar Radiation Modification (SRM)," which entails changes in the planetary radiation budget by measures such as injecting aluminum silicate particles into the stratosphere. Like CDR, there is considerable technological and scientific uncertainty regarding SRM, and it has to date received much less attention in the integrated assessment literature. Additionally, fusion power has been proposed for many decades, without yet coming to fruition. There is no way of knowing at this time whether these speculative possibilities, or others that have not even been imagined, will materialize.

# Appendix B. Regret Computations

## Table B-1

## Regrets

## $\alpha = 0.000125$, $\beta = 0.018$

## Part 1 of 3

## Parameter Combinations (Model used for policy) (Columns)

| Actual World | δ1 m1 | δ2 m1 | δ3 m1 | δ4 m1 | δ5 m1 | δ6 m1 | δ7 m1 | δ1 m2 | δ2 m2 | δ3 m2 | δ4 m2 | δ5 m2 | δ6 m2 | δ7 m2 |
|---|---|---|---|---|---|---|---|---|---|---|---|---|---|---|
| δ1 m1 | 0.000 | 0.312 | 1.015 | 1.865 | 2.735 | 3.567 | 4.340 | 0.043 | 0.071 | 0.428 | 0.962 | 1.574 | 2.208 | 2.833 |
| δ2 m1 | 0.137 | 0.000 | 0.075 | 0.225 | 0.392 | 0.551 | 0.696 | 0.253 | 0.033 | 0.006 | 0.074 | 0.182 | 0.303 | 0.424 |
| δ3 m1 | 0.239 | 0.040 | 0.000 | 0.020 | 0.059 | 0.102 | 0.142 | 0.347 | 0.110 | 0.020 | 0.000 | 0.013 | 0.039 | 0.070 |
| δ4 m1 | 0.268 | 0.076 | 0.012 | 0.000 | 0.006 | 0.018 | 0.032 | 0.361 | 0.145 | *0.049* | 0.011 | 0.000 | 0.002 | 0.009 |
| δ5 m1 | 0.264 | 0.090 | 0.025 | 0.004 | 0.000 | 0.002 | 0.006 | 0.345 | 0.152 | 0.062 | 0.023 | 0.006 | 0.001 | 0.000 |
| δ6 m1 | 0.249 | 0.093 | 0.032 | 0.009 | 0.002 | 0.000 | 0.001 | 0.319 | 0.148 | 0.066 | 0.029 | 0.011 | 0.004 | 0.001 |
| δ7 m1 | 0.230 | 0.090 | 0.034 | 0.012 | 0.004 | 0.001 | 0.000 | 0.293 | 0.139 | 0.066 | 0.031 | 0.014 | 0.006 | 0.002 |
| δ1 m2 | 0.056 | 0.761 | 1.924 | 3.238 | 4.546 | 5.778 | 6.911 | 0.000 | 0.280 | 0.955 | 1.832 | 2.787 | 3.752 | 4.689 |
| δ2 m2 | 0.046 | 0.041 | 0.255 | 0.532 | 0.808 | 1.060 | 1.283 | 0.127 | 0.000 | 0.079 | 0.251 | 0.455 | 0.662 | 0.859 |
| δ3 m2 | 0.143 | 0.005 | 0.023 | 0.093 | 0.173 | 0.248 | 0.315 | 0.236 | 0.043 | 0.000 | 0.024 | 0.075 | 0.134 | 0.192 |
| δ4 m2 | 0.193 | 0.033 | 0.000 | 0.012 | 0.038 | 0.065 | 0.090 | 0.278 | 0.086 | 0.015 | 0.000 | 0.008 | 0.026 | 0.046 |
| δ5 m2 | 0.208 | 0.054 | 0.007 | 0.000 | 0.007 | 0.017 | 0.028 | 0.283 | 0.107 | 0.032 | 0.006 | 0.000 | 0.003 | 0.010 |
| δ6 m2 | 0.207 | 0.064 | 0.015 | 0.001 | 0.000 | 0.004 | 0.008 | 0.274 | 0.113 | 0.042 | 0.013 | 0.002 | 0.000 | 0.001 |
| δ7 m2 | 0.198 | 0.068 | 0.020 | 0.004 | 0.000 | 0.000 | 0.002 | 0.258 | 0.113 | 0.047 | 0.018 | 0.006 | 0.001 | 0.000 |
| δ1 m3 | 0.086 | 0.897 | 2.186 | 3.628 | 5.066 | 6.396 | 7.627 | 0.002 | 0.351 | 1.113 | 2.083 | 3.133 | 4.188 | 5.209 |
| δ2 m3 | 0.031 | 0.063 | 0.315 | 0.627 | 0.933 | 1.210 | 1.454 | 0.104 | 0.002 | 0.110 | 0.310 | 0.541 | 0.771 | 0.989 |
| δ3 m3 | 0.123 | 0.002 | 0.036 | 0.119 | 0.210 | 0.295 | 0.369 | 0.212 | 0.032 | 0.001 | 0.037 | 0.099 | 0.167 | 0.232 |
| δ4 m3 | 0.176 | 0.025 | 0.001 | 0.019 | 0.050 | 0.081 | 0.110 | 0.258 | 0.074 | 0.010 | 0.001 | 0.014 | 0.036 | 0.059 |
| δ5 m3 | 0.195 | 0.046 | 0.004 | 0.001 | 0.011 | 0.023 | 0.035 | 0.269 | 0.096 | 0.026 | 0.003 | 0.000 | 0.006 | 0.015 |
| δ6 m3 | 0.196 | 0.058 | 0.012 | 0.001 | 0.001 | 0.006 | 0.012 | 0.263 | 0.105 | 0.037 | 0.010 | 0.001 | 0.000 | 0.003 |
| δ7 m3 | 0.190 | 0.063 | 0.017 | 0.003 | 0.000 | 0.001 | 0.004 | 0.250 | 0.106 | 0.042 | 0.015 | 0.004 | 0.000 | 0.000 |
| δ1 m4 | 0.303 | 1.691 | 3.658 | 5.783 | 7.855 | 9.784 | 11.545 | 0.073 | 0.793 | 2.022 | 3.497 | 5.053 | 6.594 | 8.073 |
| δ2 m4 | 0.004 | 0.230 | 0.687 | 1.187 | 1.654 | 2.067 | 2.426 | 0.028 | 0.061 | 0.324 | 0.678 | 1.050 | 1.407 | 1.736 |
| δ3 m4 | 0.051 | 0.017 | 0.138 | 0.295 | 0.446 | 0.578 | 0.692 | 0.117 | 0.002 | 0.040 | 0.141 | 0.258 | 0.374 | 0.480 |
| δ4 m4 | 0.104 | 0.003 | 0.022 | 0.077 | 0.136 | 0.189 | 0.235 | 0.175 | 0.027 | 0.001 | 0.024 | 0.066 | 0.110 | 0.152 |
| δ5 m4 | 0.135 | 0.017 | 0.001 | 0.019 | 0.044 | 0.068 | 0.090 | 0.201 | 0.053 | 0.005 | 0.002 | 0.015 | 0.033 | 0.052 |
| δ6 m4 | 0.149 | 0.030 | 0.001 | 0.003 | 0.014 | 0.026 | 0.037 | 0.209 | 0.068 | 0.015 | 0.001 | 0.002 | 0.009 | 0.018 |
| δ7 m4 | 0.152 | 0.039 | 0.006 | 0.000 | 0.004 | 0.010 | 0.016 | 0.207 | 0.076 | 0.023 | 0.004 | 0.000 | 0.002 | 0.006 |
| δ1 m5 | 0.364 | 1.887 | 4.012 | 6.298 | 8.521 | 10.589 | 12.474 | 0.099 | 0.907 | 2.245 | 3.839 | 5.513 | 7.168 | 8.755 |
| δ2 m5 | 0.006 | 0.278 | 0.783 | 1.327 | 1.831 | 2.277 | 2.663 | 0.018 | 0.084 | 0.383 | 0.773 | 1.178 | 1.565 | 1.920 |
| δ3 m5 | 0.040 | 0.027 | 0.168 | 0.342 | 0.507 | 0.651 | 0.773 | 0.101 | 0.001 | 0.056 | 0.171 | 0.302 | 0.429 | 0.545 |
| δ4 m5 | 0.092 | 0.001 | 0.031 | 0.095 | 0.160 | 0.219 | 0.269 | 0.159 | 0.020 | 0.002 | 0.034 | 0.082 | 0.131 | 0.178 |
| δ5 m5 | 0.124 | 0.012 | 0.003 | 0.025 | 0.054 | 0.081 | 0.105 | 0.188 | 0.045 | 0.003 | 0.004 | 0.020 | 0.042 | 0.063 |
| δ6 m5 | 0.139 | 0.025 | 0.001 | 0.005 | 0.018 | 0.032 | 0.044 | 0.198 | 0.061 | 0.012 | 0.000 | 0.004 | 0.013 | 0.023 |
| δ7 m5 | 0.144 | 0.035 | 0.004 | 0.000 | 0.005 | 0.013 | 0.019 | 0.198 | 0.070 | 0.019 | 0.003 | 0.000 | 0.003 | 0.008 |
| δ1 m6 | 0.431 | 2.099 | 4.394 | 6.851 | 9.235 | 11.449 | 13.467 | 0.130 | 1.030 | 2.486 | 4.206 | 6.006 | 7.783 | 9.485 |
| δ2 m6 | 0.012 | 0.332 | 0.889 | 1.479 | 2.024 | 2.503 | 2.918 | 0.011 | 0.111 | 0.449 | 0.877 | 1.318 | 1.736 | 2.119 |
| δ3 m6 | 0.030 | 0.040 | 0.202 | 0.394 | 0.574 | 0.730 | 0.863 | 0.085 | 0.002 | 0.074 | 0.206 | 0.350 | 0.489 | 0.615 |
| δ4 m6 | 0.079 | 0.001 | 0.042 | 0.115 | 0.187 | 0.251 | 0.306 | 0.144 | 0.014 | 0.006 | 0.045 | 0.100 | 0.155 | 0.207 |
| δ5 m6 | 0.112 | 0.009 | 0.006 | 0.033 | 0.066 | 0.096 | 0.122 | 0.175 | 0.037 | 0.001 | 0.007 | 0.027 | 0.052 | 0.076 |
| δ6 m6 | 0.129 | 0.021 | 0.000 | 0.008 | 0.023 | 0.039 | 0.053 | 0.187 | 0.054 | 0.009 | 0.000 | 0.006 | 0.017 | 0.029 |
| δ7 m6 | 0.136 | 0.030 | 0.002 | 0.001 | 0.008 | 0.016 | 0.024 | 0.189 | 0.064 | 0.016 | 0.002 | 0.001 | 0.005 | 0.011 |
| **Max Regret** | **0.431** | **2.099** | **4.394** | **6.851** | **9.235** | **11.449** | **13.467** | **0.361** | **1.030** | **2.486** | **4.206** | **6.006** | **7.783** | **9.485** |

**Table B-1 (continued)**

**Regrets**

$\alpha = 0.000125, \beta = 0.018$

**Part 2 of 3**

**Parameter Combinations (Model used for policy) (Columns)**

| Actual World | δ1 m3 | δ2 m3 | δ3 m3 | δ4 m3 | δ5 m3 | δ6 m3 | δ7 m3 | δ1 m4 | δ2 m4 | δ3 m4 | δ4m4 | δ5 m4 | δ6 m4 | δ7 m4 |
|---|---|---|---|---|---|---|---|---|---|---|---|---|---|---|
| δ1 m1 | 0.062 | 0.044 | 0.338 | 0.808 | 1.364 | 1.951 | 2.539 | 0.167 | 0.011 | 0.089 | 0.321 | 0.651 | 1.037 | 1.452 |
| δ2 m1 | 0.280 | 0.048 | 0.002 | 0.052 | 0.146 | 0.256 | 0.370 | 0.407 | 0.141 | 0.027 | 0.004 | 0.035 | 0.095 | 0.171 |
| δ3 m1 | 0.372 | 0.130 | 0.029 | 0.001 | 0.007 | 0.029 | 0.056 | 0.483 | 0.229 | 0.093 | 0.029 | 0.004 | 0.002 | 0.013 |
| δ4 m1 | 0.383 | 0.163 | 0.060 | 0.017 | 0.002 | 0.001 | 0.006 | 0.479 | 0.252 | 0.125 | 0.057 | 0.022 | 0.006 | 0.001 |
| δ5 m1 | 0.364 | 0.168 | 0.073 | 0.029 | 0.009 | 0.002 | 0.000 | 0.447 | 0.246 | 0.132 | 0.069 | 0.034 | 0.016 | 0.006 |
| δ6 m1 | 0.336 | 0.162 | 0.076 | 0.035 | 0.015 | 0.006 | 0.001 | 0.409 | 0.230 | 0.129 | 0.072 | 0.040 | 0.021 | 0.011 |
| δ7 m1 | 0.308 | 0.151 | 0.075 | 0.037 | 0.018 | 0.008 | 0.003 | 0.374 | 0.212 | 0.121 | 0.070 | 0.041 | 0.024 | 0.013 |
| δ1 m2 | 0.002 | 0.212 | 0.796 | 1.584 | 2.461 | 3.362 | 4.247 | 0.052 | 0.038 | 0.304 | 0.756 | 1.318 | 1.942 | 2.592 |
| δ2 m2 | 0.149 | 0.002 | 0.054 | 0.202 | 0.389 | 0.583 | 0.771 | 0.255 | 0.047 | 0.002 | 0.054 | 0.159 | 0.290 | 0.432 |
| δ3 m2 | 0.258 | 0.057 | 0.001 | 0.015 | 0.058 | 0.112 | 0.167 | 0.359 | 0.134 | 0.032 | 0.001 | 0.009 | 0.036 | 0.073 |
| δ4 m2 | 0.298 | 0.101 | 0.022 | 0.001 | 0.004 | 0.019 | 0.037 | 0.387 | 0.178 | 0.070 | 0.020 | 0.002 | 0.001 | 0.008 |
| δ5 m2 | 0.301 | 0.121 | 0.041 | 0.009 | 0.000 | 0.001 | 0.007 | 0.381 | 0.191 | 0.089 | 0.037 | 0.013 | 0.002 | 0.000 |
| δ6 m2 | 0.289 | 0.126 | 0.051 | 0.017 | 0.004 | 0.000 | 0.001 | 0.360 | 0.189 | 0.096 | 0.047 | 0.021 | 0.008 | 0.002 |
| δ7 m2 | 0.272 | 0.124 | 0.054 | 0.022 | 0.008 | 0.002 | 0.000 | 0.336 | 0.181 | 0.096 | 0.050 | 0.026 | 0.012 | 0.005 |
| δ1 m3 | 0.000 | 0.272 | 0.935 | 1.810 | 2.775 | 3.761 | 4.728 | 0.035 | 0.059 | 0.378 | 0.888 | 1.515 | 2.203 | 2.917 |
| δ2 m3 | 0.124 | 0.000 | 0.079 | 0.254 | 0.466 | 0.683 | 0.892 | 0.225 | 0.032 | 0.006 | 0.079 | 0.204 | 0.354 | 0.514 |
| δ3 m3 | 0.234 | 0.044 | 0.000 | 0.025 | 0.079 | 0.141 | 0.204 | 0.332 | 0.115 | 0.022 | 0.000 | 0.017 | 0.052 | 0.096 |
| δ4 m3 | 0.278 | 0.088 | 0.016 | 0.000 | 0.009 | 0.028 | 0.049 | 0.366 | 0.161 | 0.059 | 0.014 | 0.001 | 0.003 | 0.014 |
| δ5 m3 | 0.286 | 0.110 | 0.034 | 0.006 | 0.000 | 0.003 | 0.011 | 0.364 | 0.178 | 0.080 | 0.031 | 0.009 | 0.001 | 0.001 |
| δ6 m3 | 0.278 | 0.117 | 0.045 | 0.014 | 0.003 | 0.000 | 0.001 | 0.348 | 0.179 | 0.089 | 0.041 | 0.017 | 0.006 | 0.001 |
| δ7 m3 | 0.264 | 0.117 | 0.050 | 0.019 | 0.006 | 0.001 | 0.000 | 0.327 | 0.174 | 0.090 | 0.046 | 0.022 | 0.010 | 0.004 |
| δ1 m4 | 0.046 | 0.652 | 1.743 | 3.085 | 4.525 | 5.972 | 7.377 | 0.000 | 0.232 | 0.829 | 1.662 | 2.631 | 3.668 | 4.728 |
| δ2 m4 | 0.039 | 0.040 | 0.263 | 0.584 | 0.931 | 1.272 | 1.590 | 0.110 | 0.000 | 0.077 | 0.260 | 0.495 | 0.749 | 1.006 |
| δ3 m4 | 0.134 | 0.005 | 0.027 | 0.113 | 0.222 | 0.332 | 0.435 | 0.218 | 0.043 | 0.000 | 0.028 | 0.091 | 0.169 | 0.252 |
| δ4 m4 | 0.192 | 0.037 | 0.000 | 0.016 | 0.052 | 0.094 | 0.135 | 0.272 | 0.092 | 0.018 | 0.000 | 0.011 | 0.035 | 0.065 |
| δ5 m4 | 0.217 | 0.063 | 0.009 | 0.000 | 0.010 | 0.026 | 0.044 | 0.290 | 0.120 | 0.040 | 0.008 | 0.000 | 0.005 | 0.015 |
| δ6 m4 | 0.224 | 0.079 | 0.021 | 0.002 | 0.001 | 0.006 | 0.014 | 0.290 | 0.133 | 0.054 | 0.018 | 0.003 | 0.000 | 0.002 |
| δ7 m4 | 0.220 | 0.086 | 0.029 | 0.007 | 0.000 | 0.001 | 0.004 | 0.281 | 0.137 | 0.062 | 0.026 | 0.009 | 0.002 | 0.000 |
| δ1 m5 | 0.066 | 0.751 | 1.942 | 3.393 | 4.944 | 6.500 | 8.008 | 0.001 | 0.282 | 0.945 | 1.853 | 2.902 | 4.021 | 5.163 |
| δ2 m5 | 0.028 | 0.058 | 0.315 | 0.670 | 1.049 | 1.418 | 1.762 | 0.092 | 0.001 | 0.103 | 0.311 | 0.572 | 0.851 | 1.130 |
| δ3 m5 | 0.117 | 0.002 | 0.039 | 0.140 | 0.261 | 0.382 | 0.495 | 0.198 | 0.033 | 0.001 | 0.041 | 0.115 | 0.203 | 0.294 |
| δ4 m5 | 0.176 | 0.029 | 0.000 | 0.024 | 0.067 | 0.113 | 0.158 | 0.253 | 0.080 | 0.012 | 0.001 | 0.017 | 0.047 | 0.081 |
| δ5 m5 | 0.203 | 0.055 | 0.006 | 0.001 | 0.015 | 0.034 | 0.054 | 0.275 | 0.109 | 0.033 | 0.005 | 0.000 | 0.008 | 0.021 |
| δ6 m5 | 0.212 | 0.071 | 0.016 | 0.001 | 0.002 | 0.009 | 0.018 | 0.277 | 0.124 | 0.048 | 0.014 | 0.002 | 0.000 | 0.004 |
| δ7 m5 | 0.211 | 0.080 | 0.025 | 0.005 | 0.000 | 0.002 | 0.006 | 0.271 | 0.129 | 0.057 | 0.022 | 0.006 | 0.001 | 0.000 |
| δ1 m6 | 0.091 | 0.860 | 2.157 | 3.725 | 5.395 | 7.066 | 8.684 | 0.006 | 0.339 | 1.071 | 2.059 | 3.194 | 4.401 | 5.629 |
| δ2 m6 | 0.019 | 0.079 | 0.373 | 0.764 | 1.178 | 1.578 | 1.949 | 0.075 | 0.005 | 0.133 | 0.369 | 0.656 | 0.961 | 1.265 |
| δ3 m6 | 0.101 | 0.001 | 0.054 | 0.171 | 0.306 | 0.438 | 0.561 | 0.178 | 0.024 | 0.004 | 0.056 | 0.142 | 0.241 | 0.342 |
| δ4 m6 | 0.160 | 0.021 | 0.002 | 0.034 | 0.083 | 0.136 | 0.185 | 0.235 | 0.068 | 0.007 | 0.003 | 0.026 | 0.060 | 0.099 |
| δ5 m6 | 0.190 | 0.047 | 0.003 | 0.004 | 0.021 | 0.043 | 0.066 | 0.260 | 0.098 | 0.027 | 0.002 | 0.002 | 0.012 | 0.028 |
| δ6 m6 | 0.201 | 0.064 | 0.013 | 0.000 | 0.004 | 0.013 | 0.024 | 0.265 | 0.114 | 0.042 | 0.011 | 0.001 | 0.001 | 0.007 |
| δ7 m6 | 0.202 | 0.073 | 0.021 | 0.003 | 0.000 | 0.003 | 0.008 | 0.261 | 0.121 | 0.051 | 0.018 | 0.004 | 0.000 | 0.001 |
| **Max Regret** | **0.383** | **0.860** | **2.157** | **3.725** | **5.395** | **7.066** | **8.684** | **0.483** | **0.339** | **1.071** | **2.059** | **3.194** | **4.401** | **5.629** |

**Table B-1 (continued)**

**Regrets**

**$\alpha = 0.000125, \beta = 0.018$**

**Part 3 of 3**

**Parameter  Combinations (Model used for policy) (Columns), and No Abatement case**

| Actual World | δ1 m5 | δ2 m5 | δ3 m5 | δ4 m5 | δ5 m5 | δ6 m5 | δ7 m5 | δ1 m6 | δ2 m6 | δ3 m6 | δ4 m6 | δ5 m6 | δ6 m6 | δ7 m6 | No Abmt. |
|---|---|---|---|---|---|---|---|---|---|---|---|---|---|---|---|
| δ1 m1 | 0.190 | 0.017 | 0.063 | 0.258 | 0.550 | 0.899 | 1.282 | 0.214 | 0.026 | 0.044 | 0.204 | 0.459 | 0.774 | 1.124 | 16.802 |
| δ2 m1 | 0.431 | 0.163 | 0.039 | 0.004 | 0.023 | 0.074 | 0.140 | 0.456 | 0.186 | 0.053 | 0.007 | 0.015 | 0.055 | 0.113 | 2.441 |
| δ3 m1 | 0.505 | 0.250 | 0.109 | 0.038 | 0.008 | 0.001 | 0.008 | 0.526 | 0.272 | 0.127 | 0.049 | 0.013 | 0.002 | 0.004 | 0.606 |
| δ4 m1 | 0.497 | 0.270 | 0.140 | 0.068 | 0.029 | 0.010 | 0.002 | 0.516 | 0.290 | 0.156 | 0.079 | 0.037 | 0.014 | 0.004 | 0.202 |
| δ5 m1 | 0.463 | 0.262 | 0.146 | 0.079 | 0.041 | 0.020 | 0.009 | 0.480 | 0.279 | 0.160 | 0.090 | 0.049 | 0.026 | 0.012 | 0.082 |
| δ6 m1 | 0.424 | 0.245 | 0.141 | 0.081 | 0.046 | 0.026 | 0.014 | 0.439 | 0.259 | 0.153 | 0.090 | 0.053 | 0.031 | 0.017 | 0.039 |
| δ7 m1 | 0.387 | 0.225 | 0.132 | 0.078 | 0.047 | 0.028 | 0.016 | 0.400 | 0.238 | 0.143 | 0.087 | 0.053 | 0.033 | 0.020 | 0.020 |
| δ1 m2 | 0.067 | 0.023 | 0.242 | 0.639 | 1.148 | 1.721 | 2.326 | 0.084 | 0.013 | 0.189 | 0.536 | 0.993 | 1.517 | 2.077 | 24.657 |
| δ2 m2 | 0.277 | 0.061 | 0.001 | 0.037 | 0.127 | 0.245 | 0.376 | 0.298 | 0.077 | 0.004 | 0.024 | 0.099 | 0.204 | 0.324 | 3.809 |
| δ3 m2 | 0.379 | 0.152 | 0.043 | 0.004 | 0.004 | 0.026 | 0.058 | 0.399 | 0.170 | 0.055 | 0.008 | 0.002 | 0.018 | 0.045 | 1.001 |
| δ4 m2 | 0.405 | 0.194 | 0.082 | 0.027 | 0.005 | 0.000 | 0.005 | 0.423 | 0.212 | 0.095 | 0.036 | 0.009 | 0.000 | 0.003 | 0.349 |
| δ5 m2 | 0.396 | 0.206 | 0.101 | 0.045 | 0.017 | 0.005 | 0.000 | 0.412 | 0.222 | 0.113 | 0.054 | 0.023 | 0.007 | 0.001 | 0.147 |
| δ6 m2 | 0.374 | 0.203 | 0.107 | 0.054 | 0.026 | 0.011 | 0.004 | 0.388 | 0.217 | 0.118 | 0.062 | 0.031 | 0.015 | 0.006 | 0.071 |
| δ7 m2 | 0.349 | 0.194 | 0.106 | 0.058 | 0.030 | 0.015 | 0.007 | 0.362 | 0.206 | 0.116 | 0.065 | 0.036 | 0.019 | 0.010 | 0.038 |
| δ1 m3 | 0.048 | 0.039 | 0.306 | 0.758 | 1.325 | 1.959 | 2.625 | 0.062 | 0.024 | 0.243 | 0.641 | 1.153 | 1.734 | 2.352 | 26.817 |
| δ2 m3 | 0.245 | 0.044 | 0.002 | 0.057 | 0.166 | 0.303 | 0.451 | 0.266 | 0.058 | 0.001 | 0.039 | 0.133 | 0.256 | 0.393 | 4.193 |
| δ3 m3 | 0.352 | 0.132 | 0.031 | 0.001 | 0.010 | 0.039 | 0.079 | 0.371 | 0.149 | 0.042 | 0.003 | 0.005 | 0.028 | 0.063 | 1.115 |
| δ4 m3 | 0.384 | 0.177 | 0.070 | 0.020 | 0.002 | 0.001 | 0.009 | 0.401 | 0.194 | 0.082 | 0.027 | 0.005 | 0.000 | 0.006 | 0.392 |
| δ5 m3 | 0.380 | 0.193 | 0.091 | 0.038 | 0.013 | 0.002 | 0.000 | 0.396 | 0.208 | 0.103 | 0.046 | 0.017 | 0.004 | 0.000 | 0.166 |
| δ6 m3 | 0.362 | 0.193 | 0.099 | 0.048 | 0.022 | 0.008 | 0.002 | 0.376 | 0.206 | 0.110 | 0.056 | 0.027 | 0.011 | 0.004 | 0.081 |
| δ7 m3 | 0.339 | 0.186 | 0.100 | 0.053 | 0.027 | 0.013 | 0.005 | 0.352 | 0.198 | 0.110 | 0.060 | 0.032 | 0.016 | 0.007 | 0.044 |
| δ1 m4 | 0.001 | 0.181 | 0.703 | 1.453 | 2.340 | 3.302 | 4.295 | 0.005 | 0.138 | 0.591 | 1.263 | 2.072 | 2.960 | 3.888 | 38.510 |
| δ2 m4 | 0.126 | 0.001 | 0.055 | 0.214 | 0.427 | 0.664 | 0.906 | 0.143 | 0.005 | 0.037 | 0.173 | 0.364 | 0.583 | 0.811 | 6.314 |
| δ3 m4 | 0.235 | 0.055 | 0.001 | 0.018 | 0.072 | 0.143 | 0.220 | 0.253 | 0.068 | 0.004 | 0.011 | 0.056 | 0.119 | 0.191 | 1.765 |
| δ4 m4 | 0.288 | 0.106 | 0.025 | 0.001 | 0.006 | 0.027 | 0.054 | 0.304 | 0.119 | 0.033 | 0.003 | 0.003 | 0.019 | 0.043 | 0.648 |
| δ5 m4 | 0.304 | 0.133 | 0.048 | 0.012 | 0.000 | 0.002 | 0.011 | 0.319 | 0.147 | 0.058 | 0.016 | 0.002 | 0.001 | 0.007 | 0.284 |
| δ6 m4 | 0.303 | 0.145 | 0.063 | 0.023 | 0.006 | 0.000 | 0.001 | 0.316 | 0.157 | 0.072 | 0.029 | 0.009 | 0.001 | 0.000 | 0.142 |
| δ7 m4 | 0.293 | 0.148 | 0.070 | 0.031 | 0.012 | 0.003 | 0.000 | 0.305 | 0.159 | 0.079 | 0.037 | 0.015 | 0.005 | 0.001 | 0.078 |
| δ1 m5 | 0.000 | 0.224 | 0.806 | 1.625 | 2.587 | 3.626 | 4.697 | 0.001 | 0.175 | 0.683 | 1.418 | 2.297 | 3.258 | 4.258 | 41.262 |
| δ2 m5 | 0.107 | 0.000 | 0.076 | 0.259 | 0.497 | 0.757 | 1.021 | 0.123 | 0.001 | 0.054 | 0.213 | 0.427 | 0.669 | 0.918 | 6.820 |
| δ3 m5 | 0.214 | 0.043 | 0.000 | 0.028 | 0.093 | 0.174 | 0.260 | 0.231 | 0.055 | 0.001 | 0.018 | 0.074 | 0.147 | 0.227 | 1.923 |
| δ4 m5 | 0.269 | 0.093 | 0.018 | 0.000 | 0.011 | 0.037 | 0.068 | 0.285 | 0.106 | 0.025 | 0.001 | 0.007 | 0.028 | 0.056 | 0.712 |
| δ5 m5 | 0.289 | 0.122 | 0.041 | 0.008 | 0.000 | 0.005 | 0.016 | 0.304 | 0.135 | 0.050 | 0.012 | 0.000 | 0.002 | 0.011 | 0.315 |
| δ6 m5 | 0.290 | 0.135 | 0.056 | 0.019 | 0.004 | 0.000 | 0.002 | 0.304 | 0.148 | 0.065 | 0.024 | 0.006 | 0.000 | 0.001 | 0.158 |
| δ7 m5 | 0.283 | 0.140 | 0.065 | 0.027 | 0.009 | 0.002 | 0.000 | 0.295 | 0.151 | 0.073 | 0.032 | 0.012 | 0.003 | 0.000 | 0.087 |
| δ1 m6 | 0.001 | 0.273 | 0.919 | 1.812 | 2.854 | 3.975 | 5.128 | 0.000 | 0.217 | 0.783 | 1.587 | 2.540 | 3.578 | 4.655 | 44.201 |
| δ2 m6 | 0.089 | 0.001 | 0.102 | 0.310 | 0.574 | 0.859 | 1.147 | 0.104 | 0.000 | 0.075 | 0.258 | 0.497 | 0.763 | 1.035 | 7.363 |
| δ3 m6 | 0.194 | 0.033 | 0.001 | 0.041 | 0.117 | 0.209 | 0.304 | 0.210 | 0.043 | 0.000 | 0.029 | 0.095 | 0.178 | 0.268 | 2.094 |
| δ4 m6 | 0.251 | 0.080 | 0.012 | 0.001 | 0.018 | 0.049 | 0.085 | 0.266 | 0.093 | 0.018 | 0.000 | 0.012 | 0.038 | 0.071 | 0.782 |
| δ5 m6 | 0.274 | 0.110 | 0.034 | 0.005 | 0.000 | 0.008 | 0.022 | 0.288 | 0.123 | 0.042 | 0.008 | 0.000 | 0.005 | 0.017 | 0.348 |
| δ6 m6 | 0.278 | 0.126 | 0.049 | 0.015 | 0.002 | 0.000 | 0.004 | 0.291 | 0.138 | 0.058 | 0.019 | 0.004 | 0.000 | 0.002 | 0.175 |
| δ7 m6 | 0.272 | 0.132 | 0.059 | 0.023 | 0.007 | 0.001 | 0.000 | 0.284 | 0.143 | 0.067 | 0.028 | 0.010 | 0.002 | 0.000 | 0.097 |
| **Max Regret** | **0.505** | **0.273** | **0.919** | **1.812** | **2.854** | **3.975** | **5.128** | **0.526** | **0.290** | **0.783** | **1.587** | **2.540** | **3.578** | **4.655** | **44.201** |

**Figure B-1: Color-shaded plot of the entries in Table B-1**

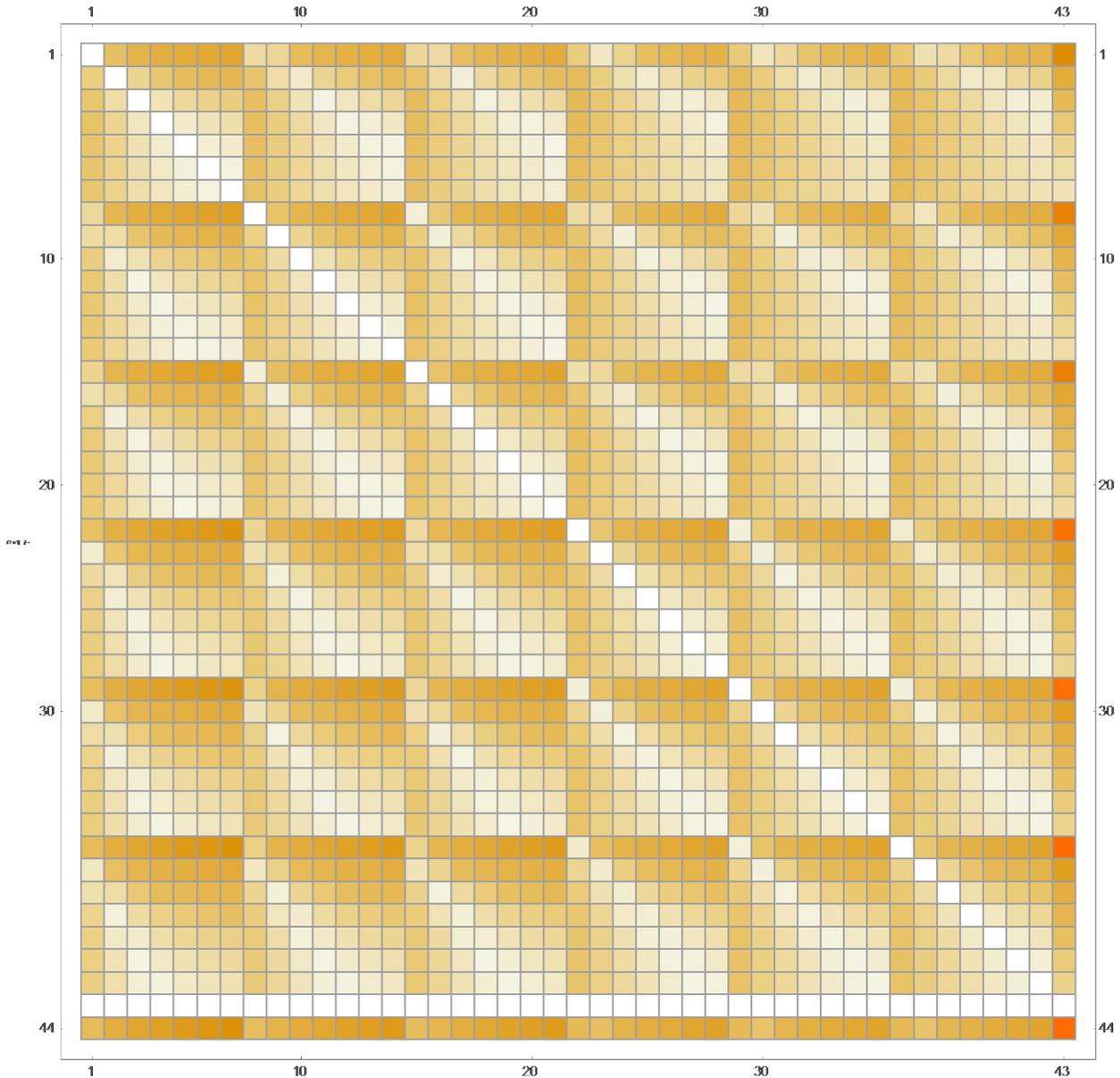

In this color plot, each cell corresponds to an entry in Table B-1. Lighter shading towards white means that the values of the cells are relatively low; darker shading towards red means the values are relatively high. The white main diagonal reflects that if the $\{\delta, m\}$ combination chosen for policy is the same as that which best describes the actual world, the regret will be zero. The highest values of the regrets are associated with the No Abatement policy. The lighter-shaded, diagonally shaped areas indicate that if the $\{\delta, m\}$ combination chosen for policy is close to the actual $\{\delta, m\}$ combination, the regret tends to be small.

# References

Ackerman, F., S. DeCanio, R. Howarth, and K. Sheeran, (2009): Limitations of integrated assessment models of climate change. *Climatic Change* **95**, 297-315.

Arrow, K., M. Cropper, C. Gollier, B. Groom, G. Heal, R. Newell, W. Nordhaus, R. Pindyck, W. Pizer, P. Portney, T. Sterner, R. Tol, and M. Weitzman, (2014): Should Governments Use a Declining Discount Rate in Project Analysis? *Review of Environmental Economics and Policy*, **8**, 145-163.

__________, (2013). Determining Benefits and Costs for Future Generations. *Science* **341**, 26 July: 349-350.

Barro, R. J., and X. Sala-i-Martin, (1995): *Economic Growth*. New York: McGraw-Hill, Inc.

Clarke, L., K. Jiang, et al., (2014). Assessing Transformation Pathways. In O. Edenhofer et al., Eds., *Climate Change 2014: Mitigation of Climate Change. Contribution of Working Group III to the Fifth Assessment Report of the Intergovernmental Panel on Climate Change.* Cambridge University Press, Cambridge, United Kingdom and New York, NY, USA.

Collins, M., et al., (2013): Long-term Climate Change: Projections, Commitments and Irreversibility. In *Climate Change 2013: The Physical Science Basis. Contribution of Working Group I to the Fifth Assessment Report of the Intergovernmental Panel on Climate Change* [Stocker, T. F., et al., Eds.]. Cambridge University Press, United Kingdom and New York, NY, USA.

Dasgupta, P., (2008): Discounting climate change. *Journal of Risk and Uncertainty* 37, 141-169.

__________, (2019): Ramsey and Intergenerational Welfare Economics, Stanford Encyclopedia of Philosophy, https://plato.stanford.edu/entries/ramsey-economics/, 1-15.

Emmerling, J., L. Drouet, K.-I. van der Wijst, D. van Vuuren, V. Bosetti, and M. Tavoni et al., (2019): The role of the discount rate for emission pathways and negative emissions. Environmental Research Letters **14** 104008.

Flato, G., et al., (2013): Evaluation of Climate Models. *Climate Change 2013: The Physical Science Basis. Contribution of Working Group I to the Fifth Assessment Report of the Intergovernmental Panel on Climate Change*, Stocker, T. F., D. Qin, G. -K. Plattner, M. Tignor, S. K. Allen, J. Boschung, A. Nauels, Y. Xia, V. Bex, and P. M. Midgeley, Eds., Cambridge Univ. Press, 741-866.

Heal, G. and A. Milner, (2014): Uncertainty and Decision Making in Climate Change Economics, *Review of Environmental Economics and Policy*, **8**, 120-137.

IPCC, (2018): Global Warming of 1.5C. An IPCC Special Report on the impacts of global warming of 1.5C above pre-industrial levels and related global greenhouse gas emission pathways, in the context of strengthening the global response to the threat of climate change, sustainable development, and efforts to eradicate poverty [Masson-Delmotte et al. Eds.].

IPCC, (2021): Climate Change 2021: *The Physical Science Basis. Contribution of Working Group I to the Sixth Assessment Report of the Intergovernmental Panel on Climate Change* [Masson-Delmotte, V., P. Zhai, A. Pirani, S.L. Connors, C. Péan, S. Berger, N. Caud, Y. Chen, L. Goldfarb, M.I. Gomis, M. Huang, K. Leitzell, E. Lonnoy, J.B.R. Matthews, T.K. Maycock, T. Waterfield, O. Yelekçi, R. Yu, and B. Zhou (Eds.)]. Cambridge University Press. In Press.

Knutti, R., (2016): The end of model democracy? *Climatic Change,* **102**, 395-404, doi:10.1007/s10584-010-9800-2.

Knutti, R., R. Furrer, C. Tebaldi, J. Cermak, and G. A. Meehl, (2010): Challenges in combining projections from multiple climate models. *J. Climate*, **23**, 2739-2758, doi:10.1175/2009JCLI3361.1.

Lee, J. Y., and Coauthors, (2021): Future Global Climate: Scenario-Based Projections and Near-Term Information. *Climate Change 2021: The Physical Science Basis. Contribution of Working Group I to the Sixth Assessment Report of the Intergovernmental Panel on Climate Change.*, Masson-Delmotte, V., P. Zhai, A. Pirani, S. L. Connors, C. Péan, S. Berger, N. Caud, Y. Chen, L. Goldfarb, M. I. Gomis, M. Huang, K. Leitzell, E. Lonnoy, J. B. R. Matthews, T. K. Maycock, T. Waterfield, O. Yelekçi, R. Yu, and B. Zhou, Eds., Cambridge Univ. Press.

Manski, C. F., A. H. Sanstad, and S. J. DeCanio, (2021): Addressing partial identification in climate modeling and policy analysis. *Proc. Natl. Acad. Sci.,* **118**, doi:10.1073/pnas.2022886118.

Matthews, H. D., N. P. Gillett, P. A. Stott, and K. Zickfeld, (2009): The proportionality of global warming to cumulative carbon emissions. *Nature,* **459**, 829-832, doi:10.1038/nature08047.

M. Meinshausen et al., (2011):  The RCP greenhouse gas concentrations and their extensions from 1765 to 2300. *Climatic Change* **109**, 213-241, DOI 10.1007/s10584-011-0156-z.

Nordhaus, W. D., (1991): To Slow or Not to Slow: The Economics of the Greenhouse Effect. *The Economic Journal* **101** (407), July: 920-937.

__________, (1994): *Managing the Global Commons: The Economics of Climate Change,* Cambridge, MA: MIT Press.

__________, (2007), "A Review of the Stern Review on the Economics of Climate Change," *Journal of Economic Literature*, **45**(3), 686–702.

__________, (2019), Climate Change: The Ultimate Challenge for Economics. *American Economic Review,* **109**, 1991-2014.

Parker, W., (2006): Understanding pluralism in climate modeling. *Found. Sci,.* **11**, 349-368, doi:10.1007/s10699-005-3196-x.

Pindyck, R., (2013): Climate Change Policy: What Do the Models Tell Us? *Journal of Economic Literature.* **51**, 860-872.

__________, (2017): The Use and Misuse of Models for Climate Policy. *Review of Environmental Economics and Policy,* 11, 100-114.

__________, (2022): *Climate Future: Averting and Adapting to Climate Change*. Oxford University Press.

Ramsey, F., (1928): A Mathematical Theory of Saving, *Economic Journal*, **38**, 543–559.

Riahi, K., et al, (2011): RCP 8.5 – A scenario of comparatively high greenhouse gas emission. *Clim. Change*, **109**, 33-57.

Sanderson, B. M., (2018): Uncertainty Quantification in Multi-Model Ensembles. Oxford Research Encyclopedia of Climate Science, Accessed October 2021, doi:10.1093/acrefore/9780190228620.013.707.

Savage, L. J., (1951): The theory of statistical decision. *J. Amer. Stat. Assoc.*, **46**, 55-67, doi:10.2307/2280094.

Smith, R. K., C. Tebaldi, D. Nychka, and L. O. Mearns, (2009): Bayesian modeling of uncertainty in ensembles of climate models. *J. Amer. Stat. Assoc.*, **104**, 97-116, doi:10.1198/jasa.2009.0007.

Stern, N., (2007): *Stern Review: the Economics of Climate Change,* Cambridge: Cambridge University Press.

Taylor, K. E., R. J. Stouffer, and G. A. Meehl, (2012): An overview of CMIP5 and the experimental design. *Bull. Amer. Meteor. Soc.,* **93**, 485-498, doi:10.1175/BAMS-D-11-00094.1.

Tebaldi, C., and R. Knutti, (2007): The use of the multi-model ensemble in probabilistic climate projections. *Philos. Trans. Roy. Soc. A,* **365**, 2053-2075, doi:10.1098/rsta.2007.2076.

U.S. Office of Management and Budget, (2003): "U.S. Office of Management and Budget Circular A-4: Regulatory Analysis." https://obamawhitehouse.archives.gov/omb/circulars_a004_a-4/.

Wald, A., (1950): *Statistical decision functions*, New York: Wiley.

Weisbach, D. and C. Sunstein, (2009): Climate Change and Discounting the Future: A Guide for the Perplexed. 27 *Yale Law and Policy Review,* 433.

Weitzman, M. L., (2001): Gamma discounting. *American Economic Review*, **91**, 260–71.

Weyant, J., (2017): Some Contributions of Integrated Assessment Models of Global Climate Change. *Review of Environmental Economics and Policy,* **11**, 115-137.

Wolfram Research, Inc., (2019), *Wolfram Mathematica 12.1.* Champaign, IL.